\documentclass[twocolumn,showpacs,titlepage,aps,prd,letter,
 tightenlines,amsmath,byrevtex,nofootinbib,floatfix]{revtex4}

\begin{document}


\title{IceCube PeV Cascade Events Initiated by Electron-Antineutrinos at Glashow Resonance}


\author{Vernon Barger}
 \email{barger@pheno.wisc.edu}
 \affiliation{Department of Physics, University of Wisconsin, Madison, WI 53706, USA}

\author{John Learned}
 \email{jgl@phys.hawaii.edu}
 \affiliation{Department of Physics and Astronomy, University of Hawaii, Honolulu, HI 96822}

\author{Sandip Pakvasa}
 \email{pakvasa@phys.hawaii.edu}
 \affiliation{Department of Physics and Astronomy, University of Hawaii, Honolulu, HI 96822}

\date{\today}

\begin{abstract}

We propose an interpretation of the two neutrino initiated cascade events with PeV energies observed by IceCube:  Ultra-high 
energy cosmic ray protons (or Fe nuclei)  scatter on CMB photons through the Delta-resonance (the Berezinsky-Zatsepin 
process)  yielding charged pions and neutrons.  The neutron decays give electron-antineutrinos which undergo neutrino 
oscillations to populate all antineutrino flavors, but the electron-antineutrino flux remains dominant.  At 6.3 PeV electron-antineutrino energy 
their annihilation on electrons in the IceCube detector is enhanced by the Glashow resonance (the W-boson) whose decays can give the PeV showers observed in the IceCube detector.  The two observed showers with  $\sim1$ PeV energies would need to be from W leptonic decays to electrons and taus.  An order of magnitude higher event rate of showers at 6.3 PeV is predicted from W to hadron decays.  This interpretation can be tested in the near term.  It has significant physics implications on the origin of the highest energy cosmic rays, since neutrino events and cosmic ray events likely share a common origin.

\end{abstract}

\pacs{14.80.Ly 12.60.Jv}

\maketitle

High energy neutrinos from the cosmos are of great interest as they should be closely related to the highest energy cosmic 
rays, whose origin is still an unsolved mystery. Experiments designed to detect these cosmogenic neutrinos \cite{BZ} have 
placed increasingly tighter upper limits on their flux\cite{AhlersNu12}. However, very recently, the IceCube (IC) 
Collaboration has reported two neutrino-induced cascade events with energies of order 1 PeV\cite{Ishihara}, which are 
likely cosmogenic. These cascade events could be i) showers from a electron, ii) showers from a tau that decays to hadrons or electrons, or iii) 
showers from the primary production of hadrons.  It has not been reported whether PeV muon-neutrino 
charged-current events have been detected as well.  The IC observations place a stringent upper bound on the overall neutrino flux. \cite{AhlersNu12}.

Studies of cosmogenic neutrinos have mostly focused on muon-neutrinos from pion, kaon, and charm 
decays\cite{Quigg},\cite{Stanev},\cite{Olinto},\cite{Yoshida},\cite{Halzen}, without any particular emphasis on the 
electron-antineutrino flux from neutron decays (except for a possible galactic electron-antineutrino source, Centaurus 
A)\cite{Weiler}.  However, electron-antineutrinos have a unique advantage because of the Glashow resonance \cite{Glashow} 
(the W-boson) at antineutrino energy of $E$= $M_W^2$/(2$m_e$) = 6.3 PeV.  This resonance gives an enhancement of 
electron-antineutrino annihilation on electrons of a factor 300 \cite{Gandhi},\cite{Xing},\cite{quant}, thus giving 
electron-antineutrinos a huge advantage over muon-neutrinos and tau-neutrinos, which can compensate for a comparatively smaller continuum flux. \cite{RajNew},\cite{StanevNew}. The W-boson decays to hadrons (68\%), to 
electrons (11\%), and to tau-leptons (11\%) will be manifest as showers in IC, on which we elaborate further below.  Our conjecture is that source of the IC PeV shower events is this resonant process.  At the Glashow resonance the earth is virtually opaque to electron antineutrinos, so the directions of such resonance associated events should be downward to horizontal. 

The primary source of cosmogenic neutrinos is cosmic ray protons (or Fe nuclei) scattering on the cosmic microwave background 
(CMB) via the Berezinsky-Zatsepin (BZ) process\cite{BZ}.
 $$ proton + \gamma(CMB) \rightarrow \Delta^+ \rightarrow \pi^+ + neutron$$
where $\Delta^+$ is the pion-nucleon resonance at $1236~MeV$. The three-body decay of the neutrons give 
electron-antineutrinos and the two-body decay of the charged pions give muon-neutrinos.  We make the assumption that the two highest energy shower events observed in the IceCube detector are W resonance events, and we discuss the implications of such a grand initiation of high 
energy neutrino astronomy.

Including the energy losses of ultra-high energy protons (with typical inverse power law energy spectra) in their propagation through the CMB, 
it is found that the generic non-resonance flux of electron-type antineutrinos has a broad maximum near 10 PeV while the flux of 
muon-type neutrinos has a broad maximum near  $1~EeV$\cite{Stanev}.  Thus, muon-neutrino events associated with the BZ process are not expected in the $1$ to $10~PeV$ range now being probed by IC. 
\\
At proton energies near $10^{19}$ eV, the cosmic ray spectrum cuts off sharply 
due to absorption over extra-galactic distances from the scattering on the CMB.  A GZK\cite{GZK}-like cut-off is seen in the 
Auger\cite{Auger} and HiRes\cite{HiRes} cosmic ray data, if one assumes that protons are the primary component of the highest energy cosmic 
rays. However, the Auger data favor a largely Fe composition.  Corresponding to the GZK cut-off for protons, the neutrino 
energy spectrum should also cut-off sharply above $10^{19}$ eV.  It has been argued that a large detector array, of order 
$1000~km^3$, or an experiment, such as ANITA\cite{Gorham}, with a huge target but higher energy threshold, would be needed to detect the BZ neutrino 
flux.\cite{Allison}

Returning to our consideration of the electron-antineutrinos from the neutron decays, neutrino oscillations will modify the 
neutrino-flavor composition ratios from an initial (1:0:0) to (2.5:1:1) \cite{flavorratios} in the tri-bi-maximal neutrino 
mixing approximation \cite{HPS}, which becomes (2-3:1:1) when recent global fits to the MNSP neutrino mixing matrix elements 
\cite{globalnu} are used.  Thus, a predominance of electron-antineutrinos from the neutron source exists even before the 
W-resonance cross section enhancement is included.  At higher neutrino energies the flavor mix is expected to be essentially 
1:1:1 from the 1:2:0 of the pion-muon decay chain\cite{PJGL}. 
Once the neutrino and antineutrino fluxes are inferred from IC observations,  the corresponding diffuse photon flux can be predicted and compared with Fermi-LAT\cite{GZKFermi} data.

The resonance events for which the W decays into hadrons will have shower energies very close to the resonance energy 6.3 PeV.
When the W decays into e, which has a branching fraction of about 11 percent, the shower energy has a  spectrum from 0 to 6.3 PeV, where y = E(e)/(6.3 PeV), with a distribution $(1-y)^2$.  The average shower energy is $6.3 PeV<y> = 1.57 PeV$.  When the W decays into tau, the showering events have a branching fraction of 11 percent times the combined branching fraction of tau to hadrons and tau to electrons, which is about 90 percent, so the net branching fraction is about 10 percent.  The corresponding shower energies are roughly
1/2 those of the two-body W to electron mode.  The Glashow resonance origin of shower events predicts the relative event rates of the various final states.  The two showers reported by IC with energies close to 1 PeV would need to be of electron or tau origin. The muons from the muonic decay of the W lose energy rapidly and will produce tracks with many mini-showers along the track and they can be easily distinguished from electron and hadron showers. 

In hadronic showers of W resonance origin, the visible event energies must be less than 6.3 PeV. These events should be dominantly W decays to two 
jets, leading to many pions and kaons.  A small fraction of the energy is lost in secondary neutrinos and the energy used in breakup of 
the nuclei.  In the IC Cherenkov detector, the energy of showering events winds down quickly to low energy particles, some of which may be 
below Cherenkov threshold.  This is in contrast to muon-neutrino initiated events, where more of the energy goes to modes 
whose Cherenkov radiation is detected but is spread over a larger distance..

In passing we note that the IC observation of cascade events bodes well for the success of the KM3 neutrino telescope in the detection of 
cosmogenic neutrinos.\cite{ANTARES}

\noindent\underline{\it Conclusions}

We have interpreted the observation of  PeV shower events in the IceCube detector as the first evidence of cosmogenic 
neutrinos. This source of the cosmic electron-antineutrinos is the BZ process in which cosmic ray protons and Fe nuclei interact with the CMB 
background photons giving neutrons that decay to electron-antineutrinos.  The cross-section for the electron-antineutrinos  interactions on 
electrons in the detector is enhanced more than two orders of magnitude by the Glashow resonance.

For a shower event originating from W decay to an electron and an electron-antineutrino, the shower energy will be between $0$ and $6.26$ PeV, with an electron energy distribution given by $(1-y)^2$, with $y = E(e)/(6.3 PeV)$.  For events from W decay to a tau, the shower energy will be roughly half that of W to e events.  

The observation of a larger sample of shower events is needed to test the Glashow resonance hypothesis.  It makes an unambiguous prediction of relative event rates from the various decay channels. Hadron shower events with energies at 6.3 PeV will occur at 10 times the rate of the lower energy showers from W to e and W to tau. 

The IC observation of shower events has significant implications for the type of cosmic neutrino 
production.  Cascade events can differentiate types of models which produce the UHE cosmic 
rays\cite{Stanev},\cite{Olinto},\cite{Auger},\cite{HiRes}.  Electron-antineutrino initiated showers could 
point to the dominance of Fe nuclei in the primaries\cite{Taylor}, as suggested by the Auger data\cite{Auger}.

If a Glashow resonance interpretation of PeV showers is confirmed by IceCube, then several matters should become the focus of present and future experiments:

First, directionality is paramount, and the extraction of directions from the cascades will be a priority and this can be done quite accurately.  The directions need to be from above the local horizon. The source 
directions should not be galactic since corresponding cosmic ray events are not seen in Extensive Air Showers 
data\cite{Auger}, and moreover the Galactic Center is not above the horizon at the South Pole.  It is not unexpected that they should come from a single or few powerful and cosmologically not distant sources, and the direction would point back to their origin.  
Gamma Ray Bursts (GRB) are a potential source of the showers.  In that case there would be a correlation with the direction of the showers and timing when a GRB is observed. Obscured GRB sources may only be seen as the showers since, unlike gamma rays, neutrinos are 
relatively unaffected by the obscuration.

Second, the higher statistics IceCube data in the analysis stage should settle whether the origin of the shower events is the Glashow resonance. The already reported two PeV shower events would be associated with W to e or W to tau decays. Moreover, W to hadrons shower events should be observed at an energy of 6.3 PeV at an order of magnitude higher rate.

Third, there must be muon events from the decay of W-bosons into a muon and a muon-antineutrino at a rate comparable to the electron shower rate and muon energy spectra the same as that of electron events.  Muons at PeV energies will 
shower heavily within the array and they should be detected as very different from cosmic ray muons originating in the 
atmosphere, which essentially never reach the array from the surface with PeV energies.

Fourth, if the IC PeV showers are electron-antineutrino Glashow resonance events, then higher energy muon-neutrino events 
should be detected (the implications are strong in this direction but not vice versa)\cite{Stanev}, but the rate is expected 
to be rather small, and even smaller if the primaries are Fe nuclei.

Fifth, experimental detector design has so far focused mainly on muon-neutrino detection, since in a 1:1:1 three-neutrino 
environment the detection of the very high energy muons wins out over contained cascades.  But with the Glashow resonance 
bump in the electron-antineutrino event rate, there is an exciting opportunity in neutrino astronomy with cascade detection.

Sixth, the discovery of the PeV shower events by IC presents an opportunity for KM3NeT\cite{KM3NeT}, which has not yet been committed in 
construction geometry, to focus on cascade detection in their design. Based on the two IC shower events in $1km^3$ years, a 
useful KM3NeT signal should be obtained in $10km^3$ years.

Seventh, the dominance of Fe in the UHE cosmic ray composition could boost the neutron
flux (and corresponding electron antineutrino flux from neutron decays) and thus enhance the rate of IceCube shower events. 

\noindent\underline{\it Acknowledgements}

We thank L.  Anchordoqui, F. Halzen, A. Karle, D. Marfatia, and T. Stanev for helpful discussions. This work was supported in part by grants from the U.S. Department of Energy. VB thanks the University of Hawaii high 
energy physics group for its hospitality.  SP thanks the Galileo Galilee Institute for Theoretical Physics for its hospitality 
and the INFN for partial support during the completion of this work.  This work was supported in part by the U.S. Department 
of Energy under grant Nos. DE-FG02- 04ER41305 and DE-FG02-04ER41291.

\nocite{*}


\begin{thebibliography}{00}
\bibitem{quant} The cross section into showers integrated over the W-resonance at 6.3 PeV is 
$4.4 \times 10^{-31} cm^2$.  The charged current cross section for muon-neutrinos is about 
$1.4 \times 10^{-33} cm^2$, so the ratio is about $300$.  See\cite{Gandhi}.
\end{thebibliography}

\end{document}